*Things in life*

# Smart, simple, sincere - why and how we should rethink connected things in our smart homes

Albrecht Kurze,
Andreas Bischof,
Arne Berger

More and more smart connected things and services turn our homes into smart environments.
They promise comfort, efficiency and security. These devices often integrate simple sensors, e.g. for temperature, light or humidity, etc. However, these smart but yet simple sensors can pose a sincere privacy risk. The sensor data enables sense-making of home attendance, domestic activities and even health conditions, often a fact that neither users nor developers are aware of or do not know how to address. Nevertheless, not all is lost or evil. This article makes a plea for how we, the ThingsCon community, might rethink smart connected things and services in our homes.
We show this in our approaches and research projects that we initiated.

The number of smart things in our homes increases. Vendors promise and users expect comfort, efficiency and security [acatech, 2023]. A networked thermostat in the living room, a motion detector at the front door or a light sensor in the hall - all these devices come with simple sensors that are tiny and integrated into new and also more and more existing household appliances. These sensors collect a wide range of data, such as temperature, humidity or movement, and can automatically adjust the heating in the home when a window is opened for ventilation, for example. However, not only cameras and microphones but also these simple sensors are a potential risk to the privacy in the home [Kurze, 2020].

In what follows, we will show how people misappropriate simple sensor data in their homes to spy on members of their household, and we will show how simple sensor data can be used for connected objects and services that foster emotional connections over distance while making eavesdropping more difficult.

In Western countries such as Germany, there has been an increase in awareness and mistrust of cameras and microphone-based devices and systems. A special case of smart technology is smart toys. In this area, products such as "Hello Barbie" and "My Friend Cayla" have attracted public attention. The latter was banned as an illegal spy device by the Federal Network Agency in Germany in 2017, resulting in them being destroyed and ownership being prohibited. However, the opportunities, dangers and privacy risks posed by simple sensors are not so well understood or prominent in public awareness.

**Why we should rethink: the danger of simple sensors**

To realize how widespread such simple sensors are, just take a walk through the nearest electronics store: refrigerators with smart thermometers that report to the smart phone when their door is open or smart door locks that record who has come and gone are now part of the standard range. Collecting simple sensor data in your own home has many advantages: it can help to save energy or make your home more comfortable and secure. But what seems harmless at first glance can also put your privacy at risk - more than many people realize.

But where exactly is the maliciousness of these sensors hidden? Simple sensors might create only thin data, but it is also big data in terms of the amount of the data collected over time [Gomez Ortega, 2022]. That data allows for sense-making of attendance at home, domestic activities and even health conditions [Kurze, 2020]. A misuse of this data easily leads to unintended implications, sometimes even to severe implications when used for surveillance purposes.

In a series of studies we gave participants from 27 households our Sensorkit with simple sensors that measure brightness, humidity, temperature and movement etc., as well as a tablet that visualizes the collected data [Kurze, 2020, 2022]. We found numerous cases of problematic uses even after short periods of time, starting with the use of data as evidence in arguments, to quantify things otherwise not objectively measurable, e.g. of wasteful behavior, the use to shame and educate others in the home, and also to surveil each other [Berger, 2023]. The 'garden example' reported in [Kurze, 2020] is such an example. One participant reported: "Well, it was quite funny because [my partner] had been out and he had somehow said: 'I've been in the garden the whole time. And then I laughed and said: 'That can't be true because the front door didn't open again until 17:30. (laughter) And then he said: 'Really? I said: 'What did you do?' - 'I think I lay on the couch for another hour and slept. I say: 'Yes, but you weren't in the garden. And then he asked: 'Have you been watching me?'

It is really astonishing that these problematic uses appear despite no really sophisticated sensors, Artificial Intelligence, cloud, external third party (Big Brother/company) and often not even an evil intent is included [Kurze, 2021b]. Instead, the user just used some learned insights to link the sensor data of a simple motion detector at the front door with knowledge about her partner - and suddenly this sensor data became an instrument of power for lateral surveillance [Richter, 2018]. In addition, the factors mentioned above (AI, cloud, 3rd parties, etc.) can potentially make the situation even worse.

The impact of simple sensors in the home does not just affect those who consciously use them. They can also affect less tech-savvy partners, children of worried parents, elderly people who have sensors installed to look after them remotely, but also anyone else who comes in and out of the home (nannies, visitors, etc.). It becomes even more noticeable when actors outside the home are involved. Landlords can monitor humidity levels in the bathroom with the best of intentions to prevent mold growth. And yet such simple sensors allow conclusions to be drawn about the type and duration of bathroom use. What's more, the processing of this data sometimes leads to misinterpretations that can have far-reaching consequences, and tracking and data analysis of supposedly simple sensors can still reveal a lot about usage behavior. And this becomes a commodity for the manufacturers themselves: one reason why consumers often have little insight into the data that a device collects about them.

This lack of transparency, not only in relation to the collected data itself, but also about where and by whom it is transmitted and how it is monitored, is at the heart of the problem. To date, there is no transparent information for consumers on this issue, for example in the form of clear information on product packaging, nor are there any safe alternatives for use. As a result, values such as convenience, efficiency and security easily come into conflict with privacy, openness and transparency. 'Well-intentioned' thus easily becomes 'accidentally evil'.

Even with all due caution, simple sensor data can also be the source for meaningful use cases and material for emotionally valuable smart objects and environments.

**How we might rethink: research approaches and desired outcomes**

We have initiated a number of research projects to find out how we - as a community of academia, practitioners, industry and the public - can surpass the status quo. The most recent projects are Simplications and Bitplush. The *Simplications* project focuses on the impact of sensors on privacy in the home as a smart environment, and the *Bitplush* project explores how smart things in the form of smart, connected plush toys can be used for closeness over distance using simple sensors in a privacy-friendly setting. Both projects are funded by the German Ministry of Education and Research (BMBF, FKZ 16KIS1868K and FKZ 16SV9117).

*Thinking beyond technology*

In technology development, especially in computer science, and computer science education, we still talk too much about technical advances and too little about the needs of people and the impacts of technologies. When designing and developing for a private space like your own home, this is neither purposeful nor responsible. In an interdisciplinary alliance of sociology, design and computer science in close cooperation with those who will be affected by this technology lies the key to do better and for success. In both projects participatory and co-design approaches help to address relevant needs and concerns right from the beginning. This includes surveys, workshops and field studies with different end-user groups (e.g. school kids, families, communal living).  Developers should discuss with future users which smart objects should be placed in the private space of the home and what they are allowed to do there: What values, wishes and needs need to be considered? Based on this, very individual smart objects can be developed. This may lead us away from technical solutions that are easily scalable and efficiently marketable. However, it opens up a space of possibility for better fitting solutions and maybe even idiosyncratic ones.

| *Uncover, understand and communicate implications* | In *Simplications* we continued the series of field studies started in a previous explorative design-driven project [Kurze, 2020]. We generated more insights of the dynamics of how unintended implications emerge and collected data for transfer in implications for use, aiming for end users, and implications for design, aiming for those that design and develop. One of the intended outcomes will be realistic reports of problematic uses actually observed [Berger, 2023] in contrast to theoretical settings or imagined fairy tales of IoT. These realistic reports should help users to understand potential risks emerging in similar situations as they are to raise awareness and to enable informed decisions. |
|---|---|
| *Co-design for creative yet privacy respecting use of sensors* | Our "Whether Bird" [Lefeuvre, 2016] is an example of a privacy respecting smart product. We developed it in a previous project together with blind and visually impaired pupils. It shows how they can help people with special needs in particular. Its name deliberately contains the English word for "whether" and not "weather". This is because it solves a problem for the young co-designers involved: They can only ever call up forecasts in their weather apps, but cannot find out what the weather was like last night, for example, or whether it is still wet on the way to school in the morning. Outside on the windowsill, a sensor measures the amount of rain that has fallen in the past few hours. A plush bird lives in the apartment, equipped with a small loudspeaker and melody generator and connected wirelessly to the sensor. The plush bird sings when you pluck its beak and does so in a slightly different way depending on whether it has rained. Only the users themselves know what each melody means. This data-frugal secret language is a way of circumventing common voice assistants for blind people, which the pupils rejected because they would be perceived as "needy and disabled" if they used them. In *Bitplush* we continue this work in the realm of smart soft toys. We created a new co-design tool, the *Wheel of Plush* for this purpose [Sontopski, 2024]. It comes with sensors and actuators integrated in plush and is intended for use in workshops. It relates to the principle of data normalization for mapping between simple inputs and outputs that still allows for private communication while potentially lowering the risks of raw data use [Stephan, 2024]. |
| | We also research the long-term appropriation of such everyday smart devices. New forms of interpersonal interaction are also made possible by simple sensor data. This is demonstrated, e.g., by the Yo-Yo Machines, which use simple, self-built tools to bring people who live physically separated from each other into contact with each other. For example, a pressure sensor in one home activates a small light in the other - and a mother knows without words that her son is sitting in his favorite armchair. It is difficult to monitor this communication and its meaning from the outside, but for mother and son it offers a simple but effective shared secret language. |
| *Education for the next generation of smart technology creators* | One promising approach is to raise the awareness and responsibility of those creating the technology in the next generation as it continues and multiplies. This could begin with the use of easy-to-use and sometimes even fun tools and methods that bring ethics into human-computer interaction and design education [Kurze, 2021a]. |

**27**

It could also be an entire course with a semester-long curriculum. In the teaching-learning project Data-I, we have initiated such a transfer of our approaches and findings from research to university teaching [Kurze, 2023]. We have integrated data-driven methods and tools such as our Sensorkit into project-based learning along the human-centered design process for interactive smart systems. Multidisciplinary teaching and learning as well as the combination of theory, hands-on experience and activities to reflect on possible implications proved to be success factors.

*Transdisciplinary exchange*

As we are not only addressing the academic world or end users, but the entire community, we have organized a series of workshops in conjunction with ThingsCon in recent years: "From simple sensors to (un-)intended implications", "From (un-)intended implications for privacy to implications for design and use" and "Plushification – Soft DIY Devices for Private Communication". These workshops helped to engage with the community and bring in perspectives from disciplines and stakeholders not otherwise represented. We would also like to summarize our findings to generalizable approaches, e.g. with a transfer of findings from a single apartment to entire smart buildings [Kurze, 2024]. We find this form of outreach beyond the boundaries of technical disciplines and academic world very enriching and therefore recommend this trans-disciplinary exchange to other researchers.

**Conclusion**

The home is our most private space, which is increasingly populated by technologies that propagate comfort, efficiency and security. Every sensor in the home, either smart or simple, and the collected data come with potentially severe risks for autonomy and privacy of those affected by these systems. However, not all is lost. We discussed our approaches for better informed users, researchers, current as well as next generation developers and designers for more innovative yet responsible smart connected products.

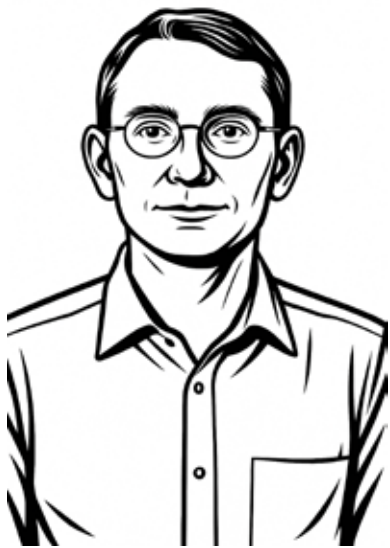 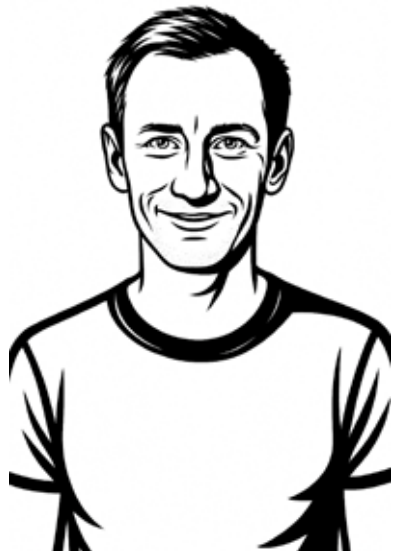 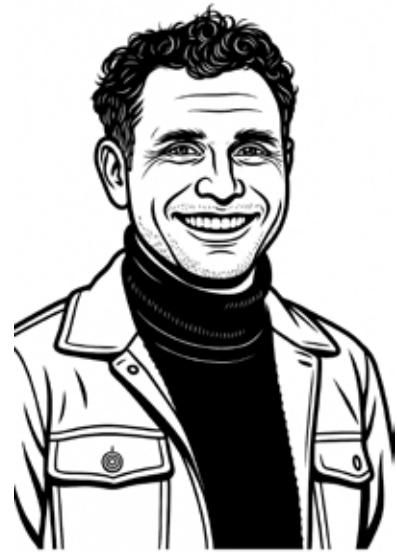

**Albrecht Kurze** is a post-doctoral researcher at the chair Media Informatics at TU Chemnitz and one of the principal investigators in Simplications and Bitplush. With a background in computer science his research interests are on the intersection of Ubiquitous HCI and human centered IoT: How do sensors, data and connectedness in smart products and environments allow for new interactions and innovation and how do we cope with the implications that they create, i.e. for privacy in the home. Albrecht has co-hosted several workshops at previous Things editions.

**Arne Berger** is a Professor of Human-Computer Interaction at Hochschule Anhalt. He is fascinated by the complex, idiosyncratic and unintended interactions between humans and digital technology. His work is influenced by the Scandinavian tradition of Participatory Design, which recognizes that those who will be affected by a future technology should have an active say in its creation. Arne's research focuses on the early phases of design and development processes, and he is particularly interested in how errors, failures, blips, and oversights shape how we think about future technology.

With a background in sociology, media communication and cultural anthropology, **Andreas Bischof** studies how technology, science and society interact with each other. Currently, he is Juniorprofessor at Chemnitz University of Technology and principal investigator in several research projects. His research focuses on the development, implication and implications of use of new technologies such as robots in health care, artificial intelligence in the workplace and smart home devices.